\documentclass[12pt]{article}
\def\be{\begin{equation}}
\def\ee{\end{equation}}
\def\ba{\begin{eqnarray}}
\def\ea{\end{eqnarray}}

\begin{document}

\baselineskip = 20pt

\title{Scattering and Bound State Green's Functions 
on a Plane via so(2,1) Lie Algebra}

\author{P. F. Borges$^1$\thanks{pborges@cefet-rj.br}, \,
H. Boschi-Filho$^2$\thanks{boschi@if.ufrj.br} \, and \,
A. N. Vaidya$^2$\thanks{vaidya@if.ufrj.br}
\\  
\\ 
\small \it
$^{1}$Centro Federal de Educa\c c\~ao 
Tecnol\'ogica Celso Suckow da Fonseca\\
\small \it Coordena\c c\~ao de F\'{\i}sica,  
Av. Maracan\~a, 229, Maracan\~a, \\
\small \it 20271-110 Rio de Janeiro, Brazil
\\
\\
\small \it
$^{2}$Instituto de F\'\i sica, 
Universidade Federal do Rio de Janeiro \\ 
\small \it Cidade Universit\'aria, Ilha do Fund\~ao, 
Caixa Postal 68528 \\
\small \it 21941-972 Rio de Janeiro, Brazil}

\bigskip

\maketitle
\begin{abstract} 
We calculate the Green's functions for the particle-vortex system, for two anyons on a plane with and without a harmonic regulator and in a uniform magnetic field. 
These Green's functions which describe scattering or bound states (depending on the specific potential in each case) are obtained exactly using an algebraic method related to the SO(2,1) Lie group. From these Green's functions we obtain the corresponding wave functions and for the bound states we also find the energy spectra.
\end{abstract}

\vfill
\pagebreak

\section{Introduction}

In this paper we study exactly solvable problems for one or two particles on a
plane bound or not by external potentials by constructing algebraically their Green's
functions. These Green's functions describe scattering or bound states, depending on
the kind of interaction in each situation. 
The algebraic method used here is based on the Schwinger representation 
\cite{Schwinger} for the inverse of an operator which is an integral representation
involving the exponential of the operator. The operator is identified with the
hamiltonian of the problem which can be written as a linear combination of the
generators of a Lie algebra. In particular, we are interested in the so(2,1) Lie
algebra which describe some well known problems as  the harmonic oscillator, the
hydrogen atom and the Morse potential \cite{Biedenharn62}-\cite{BoschiVaidya90a}. 
Some other problems with more involved potentials can also be described by this
algebra (see, for instance refs. \cite{Solomon71}-\cite{deSouzaDutra:9194}). 
Once the hamiltonian is written in terms of the so(2,1) generators one can use
Baker-Campbell-Hausdorff (BCH) formulas \cite{BCH}-\cite{Gilmore} to split the
exponential of the hamiltonian into a convenient product of the so(2,1) generators.
BCH formulas are also used to change the order of the product of the exponentials of
generators to simplify the computation of the Green's functions. This method was used
to describe the Dirac electron in a Coulomb potential \cite{MS82} and the discussion
presented here is a non-relativistic version modified to include other potentials
\cite{BoschiVaidya90a,Vaidya89}. As we will see, for the simplest cases the
hamiltonian is identified simply with just {\sl one} so(2,1) generator. In these
particular cases just one BCH formula is used.

The two dimensional problems we are going to discuss here have been studied before in
\cite{Aharonov:1959fk}-\cite{DasnieresdeVeigy:1993rx} with other approaches, although
for instance, in \cite{Jackiw90} the so(2,1) symmetry was invoked to construct the
wave function for the particle-vortex system.
The problem of particles moving on a plane is relevant to the studies
of condensed matter systems as the fractional quantum Hall effect and anyonic superconductivity \cite{Laughlin83}-\cite{Forte92}, supersymmetry \cite{Plyushchay, Rausch}, and fault-tolerant quantum computing \cite{Kitaev}. Free anyon Green's functions have been recently used to study correlation functions of anyon interferometry \cite{Gutierrez}. 

This paper is organized as follows: 
In section 2 we discuss the particle-vortex system and its so(2,1) dynamical algebra and in section 3 we use it obtain its Green's function algebraically. From this Green's function we obtain the wave functions and find a continuous energy spectrum. In section 4, we discuss the two anyon problem on a plane within a harmonic well and obtain its Green's function using the above mentioned algebraic method related to the SO(2,1) Lie group. From this Green's function we obtain the corresponding wave functions and the discrete energy spectrum. Then, in section 5, we obtain the Green's function for the problem of two anyons without any regulator from the results of the previous section. We also show that this problem is equivalent to the particle-vortex system, if one identifies the quantized flux of the particle-vortex with the anyon statistical parameter. Finally, in section 6 we obtain exactly the Green's function for two anyons in a uniform magnetic field and the corresponding wave functions and the discrete energy spectrum. In section 7 we present our conclusions.


\section{The particle-vortex system}

Let us start the discussion with the particle-magnetic vortex which is defined to be
a
two-dimensional system characterized by the Schr\"odinger equation  
\be
i\, \hbar\, \frac{\partial\Psi}{\partial t} = H \Psi\,.
\ee
The hamiltonian of interest is given by
\be
H = \frac 1{2M} \left( \vec p - \frac ec \vec A \right)^2
\ee
with an externally prescribed vector potential
\be
\vec A = \frac \Phi{2\pi r^2}\, {\bf e}_3 \wedge \vec r
\ee
where ${\bf e}_3$ is a constant unit vector perpendicular to the plane in which
$\,\vec r\,$ lies. 

The vector potential gives rise to a magnetic field
\ba
B &=& \nabla \wedge \vec A \cr
&=& \Phi \, \delta^2(\vec r)
\ea
with flux $\Phi = \int B(\vec r)\, d^2r$. 

A simple example is given by the motion of a charged particle around a magnetic flux
line when the force motion parallel to the flux line is ignored. If one solves the
two dimensional problem one may apply a boost along the direction of the flux line
and get the description of the three dimensional system. The experimental set-up was
considered in the Bohm-Aharonov effect \cite{Aharonov:1959fk}. 

Clearly, the system is invariant under two dimensional rotations whose generator is
given by the conserved angular momentum 
\be
J = \vec r \wedge \vec p
\ee
where $\,\vec p\,$ is the canonical momentum and $J$ generates the O(2) group of
rotations in  a plane. Also, since $H$ does not have an explicit time dependence it
is a constant of motion. In principle we expect to have two more constants of
motion.
In an interesting paper Jackiw \cite{Jackiw90} constructed them in explicit form as
\ba
D &=& \ t\ H - \frac 14 ({\vec r} \cdot {\vec p} + {\vec p} \cdot {\vec r})
\label{D}
\\
K &=& -\ t^2\ H + 2tD + \frac{Mr^2}{2}
\ea
where  the fact that ${\vec r} \cdot {\vec A} = 0$ in (\ref{D}) was used. 

It may be noted that $D$ and $K$ are generators of scale and conformal
transformations which change the Lagrangian by a total time derivative. One can
verify that 
\ba\label{HDH}
\left[\, H \, , \, D \, \right] &=& -\,i\,\hbar\, H \\
\label{DKK}
\left[\, D \, , \, K \, \right] &=& -\,i\,\hbar\, K \\
\label{KHD}
\left[\, K \, , \, H \, \right] &=& +\,2\, i\, \hbar\,D 
\ea
which are the commutation relations of the generators of the algebra associated with
the group SO(2,1). 

Since $J$ commutes with $H$, $D$ and $K$ the symmetry group of the system is the direct product SO(2)$\times $SO(2,1). In the following we consider the construction of the Green's function for the particle vortex system by making use of the
Baker-Campbell-Hausdorff (BCH) formulas for the exponentials of the generators of the algebra of SO(2,1) group. Since we separate the time variable we only need the form of the generators at $\, t=0$. 

The Green's function associated with the particle-vortex system satisfies the
equation
\ba
\left( i\hbar\,\frac\partial{\partial t} 
- H\right) G({\vec r}\ ,\ t\ ;\ {\vec r\,}^\prime, t^\prime)
= -\ \delta ({\vec r} - {\vec r\,}^\prime)\ \delta (t - t^\prime)\,.
\ea

Since $H$ is time independent one may write
\ba
G({\vec r}\ ,\ t\ ;\ {\vec r\,}^\prime, t^\prime)
= \frac 1{2\pi} \int \, dE\, G_E({\vec r}\ ,\ {\vec r\,}^\prime)\ 
e^{-iE(t - t^\prime)/\hbar\,}
\ea
and get 
\ba
(-H+E ) \  G_E({\vec r}\ ,\ {\vec r\,}^\prime)\ = \delta ({\vec r} - {\vec
r\,}^\prime)\,.
\ea
Then, using the Schwinger \cite{Schwinger} representation for the inverse of an operator one can write the Green's function as
\ba
 G_E({\vec r}\ ,\ {\vec r\,}^\prime)\ = \frac i\hbar\, \int_0^\infty \,ds\, e^{\,
i\,
(E-H + i\epsilon )\, s/\hbar\,}\  
\delta ({\vec r} - {\vec r\,}^\prime) \,,
\ea
where $s$ is usually known as the Schwinger's proper time and  $\epsilon>0$ is included to assure the convergence of the above integral.  
Let us now calculate explicitly this Green's function algebraically. 



\section{Green's  function for the particle-vortex system}



Since $H$ is invariant under rotations we may use the result
\ba
\delta ({\vec r} - {\vec r\,}^\prime)\, 
= \frac 1r \, \delta ({ r} - { r\ }^\prime)\, 
\delta ( \phi - \phi\,^\prime)\,,
\ea
where 
\ba
\delta ( \phi - \phi\,^\prime)\,
= \frac 1 {2\pi}\,  \sum_m e^{im( \phi - \phi\,^\prime)}\,,
\ea
with integer $\,m\,$. Thus if we write
\ba
G_E({\vec r}\ ,\ {\vec r\,}^\prime)\ 
= \frac 1 {2\pi}\,  \sum_m e^{im( \phi - \phi\,^\prime)}\, 
G_{E\, m}({r}\ ,\ {r\,}^\prime)\ 
\label{GErr'}
\ea
then the one dimensional Green's function is 
\ba\label{Gem1rr'}
G_{E\, m}({r}\ ,\ {r\,}^\prime)\ 
= \frac i{\hbar\, r\,^\prime}\ \int_0^\infty e^{\, i\, (E-H_m)\, s/\hbar}\  \delta
({r} - {r\,}^\prime)\,ds \,,
\ea
where
\ba
H_m = \, - \,\frac {\hbar^2}{2M}\left(  \frac{\partial^2}{\partial r^2} \, + \,
\frac
1r \frac {\partial}{\partial r} \,
- \, \frac {(m-\nu)^2}{r^2}\, \right)
\label{Hm}\,
\ea
is the radial Hamiltonian dependent on the integer angular momentum quantum number
$m$ and $\nu $ is the quantized flux:
\ba
\label{nu}
\nu = \frac {e\Phi}{2\pi \hbar c}\,.
\ea

In the following we use the set of differential operators given by 
\ba\label{T1}
T_1 & = &   \frac{\partial^2}{\partial r^2} \, + \, \frac 1r \frac
{\partial}{\partial r} \, -\, \frac {(m-\nu)^2}{r^2}
\\
T_2 &=& - \, \frac {i}{2} \left(\, r \frac \partial{\partial r} \, + \, 1 \, \right)
\label{T2}
\\
T_3 &=& -\, \frac {1}{8}\,r^2\,.\label{T3}
\ea
\noindent which can be easily related to the ones defined in the previous section.
 These operators  satisfy the $so(2,1)$ Lie algebra
\ba
\label{T121}
\left[\, T_1 \, , \, T_2 \, \right] &=& - \;\; i \,  T_1 \\
\left[\, T_2 \, , \, T_3 \, \right] &=& - \;\; i \,  T_3 \\
\left[\, T_3 \, , \, T_1 \, \right] &=& + \;\; i \, T_2 \,,
\label{T312}
\ea
as well as the operators $H, K, D$. Next we note the following representation for the
delta function
\ba
 \delta ({r} - {r\,}^\prime)\, 
= \frac {M}{4\pi i\,{r\,^\prime}^{\delta-1}}\, \int_{-i\infty}^{i\infty} 
\, e^{\frac {1}{4}\,q\,M\,({r}^2 - {r\,}^{\prime 2})} 
\, r^\delta \, dq\,,
\ea
where $r\,,\, r\,^\prime\,\ge\, 0\,$, and the arbitrary parameter $\,\delta\,$ will
be fixed later. Thus the one dimensional Green's function can be written as 
\ba
G_{E\, m}({r}\ ,\ {r\,}^\prime)\ 
= \frac {M}{4\pi\hbar\,{r\,^\prime}^{\delta}}\  \int_0^\infty e^{\, i\, E\,
s/\hbar\,}\  ds \,
 \int_{-i\infty}^{i\infty} \, e^{-\,\, \frac {qM}{4} {{{r\,}^\prime}^2}}\, 
e^{\,\frac {is\hbar}{2M}\,T_1\,}\, e^{- 2 M q\,T_3}\, 
\, r^\delta \, dq\,.
\ea

Further, one can show that the following identity holds
\ba
 e^{\,\frac{i\,s\,\hbar}{2M}\, T_1\,}\, e^{-2Mq\,T_3}\, 
=   e^{-\,i\,\zeta_3\,T_3}\,  e^{-\,i\,\zeta_2\,T_2} \, e^{-\,i\,\zeta_1\,T_1}\, 
\label{BCH1}
\ea
where  
\ba
 e^{\,\zeta_2/2} &=&  \, \frac {s\hbar}{2\,i}\, \left(\, q\, + \frac
{2i\,}{s\hbar}\,
\right)
\\
-\,i\,\zeta_3 &=&  -2M  \left({\frac {2i}{s\hbar}\, +\, 
\frac {4}{(s\hbar)^2\left(\, q\, + \displaystyle{\frac {2i\,}{s\hbar}}\,
\right)}}\right)\;.
\ea

The proof of the above relations is given, {\sl e. g.}, in 
\cite{BoschiVaidya90a,Vaidya89}
where $\zeta_1$ is also calculated. Note however that the value of $\zeta_1$ is not
needed here since we will choose $\delta$ such that 
\ba
T_1\,r^\delta=0
\label{T10}
\ea 
which implies 
\be
\delta\  =\   |\, m\, - \, \nu\, |\,.
\ee
In fact the condition (\ref{T10}) also allows $\delta=-|m-\nu|$, but these negative
values lead to unphysical solutions as will be seen in the following. 
Next, it is easy to verify that 
\ba
e^{-\,i\,b\,T_2}\, f(r)\  = \ e^{-\,b/2}\, f(r\, e^{-\,b/2})
\ea
so that
\ba
G_{E\, m}({r}\ ,\ {r\,}^\prime)
&=& \frac {M\,{r}^{\delta}}{4\pi\,\hbar\,{r\,^\prime}^{\delta}}\  \int_0^\infty \,
ds
\ \exp\left\{\frac{\, i\, E\, s}{\hbar\,}\right\}
\exp\left\{\frac{iM}{2s\hbar}\,r^2\right\}\, 
\int_{-i\infty}^{i\infty} \, dq\exp\left\{-\, \frac{qM}{4}{{r\,}^\prime}^2\right\}\,
\cr&\times& 
\exp\left\{\frac{Mr^2}{\displaystyle (s\hbar)^2(q+\frac{2i}{s\hbar})}\right\}\, 
\,
\left[\,\frac{s\hbar\,}{2i}\left(q+\frac{2i}{s\hbar\,}\right)\right]^{-(1+\delta)}.
\ea

Expanding the exponential as a power series and doing the integrations by repeatedly
using the result
\ba
\frac 1{2\pi i}\int_{-i\infty}^{i\infty} \, dq\ \frac{\exp\{-\,
\frac{qM}{4}{{r\,}^\prime}^2\}}
{\left(q+\frac{2i}{s\hbar\,}\right)^{\xi+1}} \ 
= \ 
\frac{\exp\{\, \frac {iM}{2s\hbar} {{r\,}^\prime}^2\}\, \left(-\, \frac M4
{{r\,}^\prime}^2\right)^\xi}
{\Gamma(\xi+1)} \,,
\ea
we find that the one dimensional Green's function can be written as 
\ba
G_{E\, m}({r}\ ,\ {r\,}^\prime)\ 
= \ -\, \frac {M}{\hbar}\, \  \int_0^\infty \, \frac {ds}{s} \ e^{\, i\, E\,
s/\hbar\,}\ \exp\{\frac{iM}{2s\hbar}\,(r^2\,+{{r\,}^\prime}^2)\}\, 
 e^{-\, \frac {i\pi}{2}\delta}\,
J_\delta\left(\frac{M\,r\,{r\,}^\prime}{s\hbar}\right)\,,\label{GEm}
\ea
where we have used the definition of Bessel functions 
\ba
J_\delta(z) \ = \ \left(\frac z2\right)^\delta \sum_n
\frac{\left(-\,z^2/4\right)^n}{n!\,\Gamma(n+\delta+1)}\,.
\ea

Next using the result \cite{GR}
\ba
\int_0^\infty \, dz \, e^{-\xi z} \, J_\delta(2\beta\sqrt{z})\,
J_\delta(2\gamma\sqrt{z})\
= 
\ \frac 1\xi \, e^{-\,\frac {1}{\xi}(\beta^2+\gamma^2)}\, 
I_\delta\left(\frac{2\beta\gamma}{\xi}\right) 
\ea
valid for $\Re (\delta) > -1$ and identifying  $\, I_\delta(z)\,=\, i^{-\delta}\,
J_\delta(iz) \,$ we get 
\ba\label{Gemrr'}
G_{E\, m}({r}\ ,\ {r\,}^\prime)\ 
= \  \int_0^\infty \, dE^{\,\prime} \, 
\frac {\;\; {\cal U}^{\,m}_{E^{\,\prime}}(r)\ {\cal
U}^{\,m}_{E^{\,\prime}}({r\,}^\prime)\;}
{\,E\,-\,E^{\,\prime}\,+i\,\epsilon\,}\,,
\ea
where 
\ba\label{UmEr'}
\,{\cal U}^{\,m}_{E'}(r)\ =\ \frac{\sqrt{M\,}}{\hbar\,}\ J_\delta(\sqrt{2ME'\,}\frac
r\hbar\,)\,.
\ea

Substituting this result into equation (\ref{GErr'}) we get
\ba\label{GEuu'}
G_E({\vec r}\ ,\ {\vec r\,}^\prime)\ 
= \ \int_0^\infty \, dE^{\,\prime} \, \frac
{1}{\,E\,-\,E^{\,\prime}\,+i\,\epsilon\,}\ 
 \sum_m \frac {e^{\,i\,m\,( \phi - \phi\,^\prime)}}{2\pi} \ 
{\,{\cal U}^{\,m}_{E^{\,\prime}}(r)\ {\cal U}^{\,m}_{E^{\,\prime}}({r\,}^\prime)\,}
\ea
Note that eq. (\ref{Gemrr'}) is the spectral representation of the one dimensional
Green's function (\ref{Gem1rr'}),  ${\cal U}^{\,m}_{E'}(r)$ are the corresponding one
dimensional wave functions and the energy spectrum is continuous for energy $\,E>0\,$
and angular momentum $\,m\,$, in agreement with \cite{Jackiw90}. Note that this
should be the case since the Hamiltonian (\ref{Hm}) corresponds to a particle in a
nonconfining potential of a centrifugal barrier $1/r^2$. 

It is interesting to note that the above calculation leading to the wave functions
(\ref{UmEr'}) is rather different from Jackiw's \cite{Jackiw90} group theoretical
discussion. To trace the main differences first we mention that he considered the
operators 
\begin{eqnarray}
R &=& \frac 12 \left( \frac 1a K + a H \right)
\\
S &=& \frac 12 \left( \frac 1a K - a H \right)
\end{eqnarray}
where $a$ is a fixed parameter with time dimensionality. The operators $R$, $S$ and
$D$ also close the so(2,1) Lie algebra, analogous to (\ref{HDH})-(\ref{KHD}). Then,
he calculated the eigenstates of $R$. Note that for any fixed time, as for instance
$t=0$, the operator 
\begin{equation}\label{Rjac}
R = \frac 12 \left( \frac 1a \frac{Mr^2}{2} + a H \right)
\end{equation}
has a {\sl discrete} spectrum because of the presence of the bounding potential
$r^2$. Next, using further group theoretical methods he expressed the {\sl
continuous}  eigenstates of $H$ in terms of the eigenstates of $R$. This procedure
can be understood as an infrared cutoff regularization of the continuous
eigenstates of $H$. Our results are in agreement with those present by Jackiw
\cite{Jackiw90}.
In next section we are going to discuss the problem of anyons in a harmonic well
which, as we will discuss, can be interpreted as the particle-vortex problem with a
harmonic regulator.

Before we move to next section, let us comment that the Green's function
(\ref{GEuu'}) can also be written in terms of associated Laguerre's polynomials
$L_n^\xi (x)$. Using  the relation  
\ba
J_\xi(2\sqrt{xz}) = e^{-z} (xz)^{\xi/2} 
\sum_{n=0}^\infty \frac{z^n L_n^\xi (x)}{\Gamma(n+\xi+1)}
\ea
valid for $\xi>-1$. Identifying $x=Mr^2/\hbar\,$, $z=E/2\hbar\,$ 
 we get 
\ba
\,{\cal U}^{\,m}_{E'}(r)\ =\ \frac{\sqrt{M\,}}{\hbar\,}\
e^{-E'/2\hbar\,}\,\left(\frac{E'Mr^2}{2\hbar^2}\right)^{\delta/2}
\sum_{n=0}^\infty \left(\frac{E'}{2\hbar}\right)^n 
\frac{L_n^\delta(\frac{Mr^2}{\hbar})}{\Gamma(n+\delta+1)}\,, 
\ea
so that 
\ba
G_E({\vec r}\ ,\ {\vec r\,}^\prime)\ 
= \ \frac{{M\,}}{2\pi\hbar} \int_0^\infty \, 
\frac {dE^{\,\prime}e^{-E^{\,\prime}/\hbar}}{\,E\,-\,E^{\,\prime}\,+i\,\epsilon\,}\ 
\sum_m e^{\,i\,m\,( \phi - \phi\,^\prime)}\,
\left(\frac{E^{\,\prime}M r r^{\,\prime}}{2\hbar^2}\right)^{\delta} \, 
\cr\cr
\times\,\sum_{n=0}^\infty \left(\frac{E'}{2\hbar}\right)^n 
\frac{  L_n^\delta (\frac{Mr^2}{\hbar})}{\Gamma(n+\delta+1)}
\ \,
\sum_{\ell=0}^\infty \left(\frac{E'}{2\hbar}\right)^\ell 
\frac{ L_\ell^\delta (\frac{M{r^{\,\prime}}^2}{\hbar})}{\Gamma(\ell+\delta+1)}\;.
\ea

Further, one can also perform the integration in $E'$ so that this Green's function
can be rewritten as  
\ba
G_E({\vec r}\ ,\ {\vec r\,}^\prime)\ 
&=& - \frac {M}{2\pi\hbar}\,  e^{-E/\hbar}
\,  \, 
\sum_m e^{\,i\,m\,( \phi - \phi\,^\prime)}\, 
 \left(\frac{{M r r^{\,\prime}}}{\hbar}\right)^{\delta}
\sum_{n=0}^\infty  \sum_{\ell=0}^\infty 
L_n^\delta \left(\frac{Mr^2}{\hbar}\right) 
L_\ell^\delta \left(\frac{M{r'}^2}{\hbar}\right) 
\cr\cr
&&\times\,
\,(-E)^{n+\ell+\delta}\, 
\frac{\Gamma(n+\ell+\delta+1)}{\Gamma(n+\delta+1)\Gamma(\ell+\delta+1)}
\,\Gamma\left(-n-\ell-\delta\, ,\, -\frac E\hbar\right)
\ea
where $\Gamma(a,x)$ is the incomplete gamma function.


\bigskip


\section{Two anyons in a harmonic well}
\label{Ahw}

Anyons are quasi-particles that obey  fractional statistics, i. e., an intermediate
statistics between the Bose-Einstein and Fermi-Dirac cases 
\cite{Leinaas77}-\cite{MacKenzie:1988ft}. 
The boson and fermion wave functions differ by the interchange of two or more
identical 
particles. The bosonic  wave function is completely symmetric while the fermionic is
completely antisymmetric. Then, the two-particle wave functions before and after the
interchange of two particles
are related by:
\begin{equation}\label{contour}
\psi(\vec{r}_2 , \vec{r}_1)=e^{i\theta}\, \psi(\vec{r}_1 , \vec{r}_2)
\end{equation}
where $\theta$ determines the statistics of the system. If $\theta=0$ (modulo
$2\pi$)
the system obeys Bose-Einstein statistics and if $\theta=\pi$ (modulo $2\pi$) the
system obeys Fermi-Dirac statistics. In three spatial dimensions these are the only
allowed possibilities. However, if the particles are restricted to live in two
spatial dimensions $\theta$ can assume any real value interpolating the BE and FD
statistics.

Considering the path integral formulation of quantum mechanics, the
transition amplitude between two states is proportional to $\exp\{
iS/\hbar\}$, where $S$ is the classical action. Then, to reproduce
the above behavior we consider that a two anyon system (any $\theta$) can be
represented by a conventional lagrangean $L$ plus a topological term \cite{Forte92}:
\begin{equation}
L \rightarrow L_{\theta}=L+{\theta \over \pi}\dot\phi
\end{equation}

\noindent where $\phi$ is the relative angle between $\vec{r}_1$ and $\vec{r}_2$.
This way, turning around one particle in respect to the other by an angle $\phi=\pi$
we obtain a phase:
\begin{eqnarray}
\exp\left\{{i\theta \over \pi}\int_0^{\pi}d\phi\right\} \;=\;
\exp\left\{i{\alpha\pi}\right\}\mbox{,}
\end{eqnarray}

\noindent where 
\begin{equation}
\label{alpha}
\alpha={\theta \over \pi}\;,
\end{equation}
is the statistical parameter of the two anyon system.

Here, in this section we consider a two anyon system characterized by the coordinates
$\vec{r}_1$ and $\vec{r}_2$ moving on a plane subjected to a harmonic regulator
$V({\vec{r}}_1 , {\vec{r}}_2)={1 \over 2}m_{0}\omega^2(r_1^2+r_2^2)$, where 
$m_0$ is the mass of each anyon. The use of a potential as a regulator is not
mandatory but it is usual in the literature \cite{MacKenzie:1988ft} since it
simplifies the discussion once the spectrum becomes discrete. An alternative
regulator procedure is to use boundary conditions as considered in Arovas et al
\cite{Arovas85} to calculate the second virial coefficient of the two anyon system.
Nonetheless, it is also possible to avoid the use of any regulator and consider the
case of ``free'' anyons, as will be discussed in the next section. This situation is
in fact related to the case of the magnetic vortex discussed in the previous section.
The connection of these cases will be discussed in the next section.

Introducing the center of mass and relative coordinates
$\vec{R}={\vec{r}}_1+{\vec{r}}_2$, and 
$\vec{r}={\vec{r}}_1-{\vec{r}}_2=(r,\phi)$, respectively, the lagrangean for the two
anyon system with a harmonic regulator can be written as:
\begin{equation}
L_{\alpha}={1 \over 2}M{\dot{R}}^2+{1 \over 2}M\omega^{2}R^2+{1 \over 2}
\mu({\dot{r}}^2+r^2{\dot{\phi}}^2+\omega^{2}r^2)+\alpha\hbar{\dot{\phi}}
\mbox{,}
\end{equation}

\noindent where $M=2m_0$, $\mu={m_0/2}$. The motion of the center of mass is
described by the first part of the lagrangean corresponding to a two dimensional
harmonic motion that does not contribute to the statistical behavior. From now on we
will only consider the relative motion of the two anyon system. The canonical
momenta are then given by:
\begin{equation}
p_{\phi}={\partial L \over \partial {\dot{\phi}}}=
\mu r^2\dot{\phi}+\alpha\hbar; \;\;\;\;\;\;\;\;\;\;\;\;\;\; 
p_r=\mu\dot{r}\mbox{,}
\end{equation}

\noindent and the Hamiltonian of the relative motion of the two particles is
\begin{equation}\label{M5A}
H=-{\hbar^2 \over 2\mu}
\left\lbrack{1 \over r}{\partial \over \partial r}r{\partial \over \partial r}
+{1 \over r^2}\left({\partial \over \partial\phi}-i\alpha\right)^2\right\rbrack
+{1 \over 2}\mu\omega^{2}r^2\mbox{.}
\end{equation}

Here we are going to construct the Green's function for this problem using an
algebraic method associated with the dynamical SO(2,1) Lie group 
\cite{BoschiVaidya90a,Vaidya89}.
The technique used here is a generalization of the one presented in the previous
section, and we will recover that results in the following sections. The Green's
functions for this problem satisfies the equation 
\begin{equation}\label{A2}
(H-E)G_{E}(r, r^\prime , \phi, \phi^\prime )
={\delta(r-r^\prime ) \over r}\delta(\phi-\phi^\prime )
\end{equation}
\noindent where $G_E(\vec{r}, {\vec{r}\,}^\prime )\equiv 
G_{E}(r, r^\prime , \phi, \phi^\prime )$. Decomposing this Green's function as we did
in the previous section, eq. (\ref{GErr'}), one can obtain the radial hamiltonian
$H_r$ for this problem and we write the resolvent operator as:
\begin{equation}\label{A1}
\Lambda_{r}=(H_{r}-E)=g_0+g_{1}T_{1}(r)+g_{3}T_{3}(r)
\end{equation}
where the operators $T_i$ are given by eqs. (\ref{T1})-(\ref{T3}) which satisfy the 
so(2,1) Lie algebra eqs. (\ref{T121}-\ref{T312}) and the parameters $g_i$ are given
by 
\begin{eqnarray}
g_0=-E\;\mbox{;}\;\;\;\;g_1=-{\hbar^2 \over
2\mu}\;\mbox{;}\;\;\;\;g_3=-4\mu\omega^2\;\mbox{.}
\label{gis}
\end{eqnarray}

Note that $H_r=H_m +\frac 12 \mu \omega^2 r^2$, with $H_m$ given by eq. (\ref{Hm}), if we further identify $\alpha$ with $\nu$. Using the Schwinger \cite{Schwinger}  representation  as before  we find:
\begin{equation}\label{A5}
G_{E}(r, r^\prime
)=i\int_{0}^{\infty}ds\exp[-is(g_0+g_{1}T_{1}(r)+g_{3}T_{3}(r)-i\epsilon)]{\delta(r-r
^\prime
) \over r}
\end{equation}
In addition to eq. (\ref{BCH1}), here we need another Baker-Campbell-Hausdorff
formula 
\begin{equation}
\exp\left\{-i{s \over
\hbar}(g_{1}T_{1}+g_{3}T_{3})\right\}=\exp\{-iaT_{3}\}\exp\{-ibT_{2}\}\exp\{-icT_{1}\}
\end{equation}
where the parameters $a$, $b$, $c$, and $k$ are given by:
\begin{eqnarray}
a=2{k \over g_1}\tan{k{s \over \hbar}}\mbox{,}\;\;\;\;\;b=2\ln\left(\cos{k{s \over
\hbar}}\right)\mbox{,}
\;\;\;\;\;\;c={g_1 \over k}\tan{k{s \over \hbar}}\qquad 
k=\sqrt{{g_{1}g_{3} \over 2}}\;.
\end{eqnarray}

Following ref. \cite{BoschiVaidya90a} we find 
\begin{eqnarray}\label{A6}
& &\exp\left\{-i{s \over \hbar}( g_1 T_{1}+ g_{3}T_{3})\right\}{\delta(r-r^\prime )
\over r}\nonumber\\
&&=-\,{ik\exp\{2i\pi\delta\} \over 2 g_{1}\sin(ks/\hbar)}\; r^\prime \;
\exp\left\{-{ik\over 4{g_1}}({{r^{\prime}}^{2}}+r^{2})\cot{k{s \over \hbar}}\right\}
I_{\delta}\left(-{k \,r^\prime\, r \over 2 i g_1\sin(ks/\hbar)}\right)
\end{eqnarray}

\noindent with  $\delta$ given by:
\begin{equation}\label{A7}
\delta=|m-\alpha|\;.
\end{equation}
As before, $\delta=-|m-\alpha|$ would imply non-normalizable solutions and we will
not consider this case.

Then we obtain the Green's function as:
\begin{eqnarray}\label{GErr'A}
G_{E}(r,r^{\prime})&=&\,-\,{i \over \hbar}\int_{0}^{\infty}ds
\exp\left\{-i{s \over \hbar}(g_{0}-i\epsilon)\right\}
\; {ik\exp{(2i\pi\delta)} \over 2 g_1\sin(ks/\hbar)}\; 
(\,r\,r^{\prime})^{1 \over 2}\nonumber\\
&\times&\exp\left\{-{ik\over 4 g_1}(r^{2}+{r^{\prime}}^{2})\cot(ks/\hbar)\right\}
I_{\delta}\left\lbrack{- \, k\, r\, r^{\prime} \over 2ig_{1}\sin(ks/\hbar)}
\right\rbrack
\end{eqnarray}

\noindent where  $I_{\delta}$ is the modified Bessel function of order $\delta$. 
This Bessel function is related to the associated Laguerre's polynomials 
$L_{n}^{\delta}$ by:
\begin{eqnarray}\label{A8}
I_{\delta}\left(2{\sqrt{y^\prime yz} \over 1-z}\right)
\exp\left\{-z{y^\prime +y \over 1-z}\right\}
=(y^\prime yz)^{\delta/2}(1-z)
\sum_{n=0}^{+\infty}{n! \over \Gamma(n+\delta+1)}
L_{n}^{\delta}(y)L_{n}^{\delta}(y^\prime )z^{n}\,.
\end{eqnarray}

\noindent Then integrating over the proper time $s$ one finds for the Green's
function of two anyons on a plane with a harmonic regulator: 
\begin{eqnarray}\label{fgreen}
G_{E}(r,r^\prime,\phi,\phi^\prime)&=&-\frac{1}{\pi}\sum_{m=-\infty}^{+\infty}e^{2\pi
i|m-\alpha|}\left({\mu\omega \over \hbar}\right)^{1+|m-\alpha|}
(rr^{\prime})^{|m-\alpha|}\nonumber\\
&\times&e^{im(\phi-\phi^\prime)}\sum_{n=0}^{\infty}{n!L_{n}^{|m-\alpha|}
\left({\mu\omega \over \hbar}r^2\right)L_{n}^{|m-\alpha|}
\left({\mu\omega \over \hbar}(r^\prime)^2\right) \over
\Gamma(n+|m-\alpha|+1)}\nonumber\\
&\times& \exp \left\{ -{\mu\omega \over 2\hbar}(r^2+{r^\prime}^2)\right\}
{1 \over E-\hbar\omega (2n+|m-\alpha|+1)}\;.
\end{eqnarray}

If one writes the spectral representation for the Green's function as:\footnote{Note
that the obtained Green's function is real up to a complex phase $i\pi|m-\alpha|$.
Then we are using a nonstandard definition for the spectral decomposition to
preserve
this nontrivial phase which is related to the statistics of the system.}

\begin{equation}
G(r,r^{\prime},\phi,\phi^{\prime})=\sum_{n=0}^{\infty}\sum_{m=-\infty}^{\infty}
{\psi_{n,m}(r,\phi)\psi_{n,m}(r^{\prime},\phi^{\prime}) \over E-E_{n,m}}\mbox{,}
\end{equation}
 
\noindent so that the wave functions are given by its residues 
\begin{eqnarray}\label{wavef}
\psi_{n,m}^{\alpha}(r,\phi)&=&\frac{i}{\sqrt{\pi}}e^{\pi i|m-\alpha|}e^{im\phi}
\left({\mu\omega \over \hbar}\right)^{{1 \over 2}(1+|m-\alpha|)}
r^{|m-\alpha|}\nonumber\\
&\times&{\sqrt{n!}L_{n}^{|m-\alpha|}
\left({\mu\omega \over \hbar}r^2\right) \over \sqrt{\Gamma(n+|m-\alpha|+1)}}
\exp \left\{ -{\mu\omega \over 2\hbar}r^2\right\}\mbox{.}
\end{eqnarray}

\noindent and the poles correspond to the energy spectrum
\begin{equation}\label{boundE}
E_{nm}^{\alpha}=\hbar\omega (2n+|m-\alpha|+1)
\end{equation}

\noindent where $n=0,1,2,3,...$ is the radial quantum number. This spectrum
corresponds to that of a two dimensional harmonic oscillator with angular momentum
$\vert m-\alpha \vert$. 

If we had started with particles identified with bosons then the allowed angular
momentum values would be $m=0,\pm 2,\pm 4,...$ . Had we started with fermions then the
values of the angular momentum should be $m=\pm 1,\pm 3,\pm 5,...$ . The quantized
energy levels are periodic functions of the statistical parameter $\alpha$ with
period 2, although the energy of a single state with quantum numbers $(n,m)$ is not
periodic.
These conclusions are in agreement with \cite{Arovas85,MacKenzie:1988ft} where the
second virial
coefficient for anyons has been calculated without a harmonic regulator, but
considering boundary conditions on the wave functions.


\section{Two ``free'' anyons}


In this section we are going to obtain the Green's function for two anyons as
discussed above but without any regulator. Here, the relative motion Hamiltonian
after separating the angular variable $\phi$ is 
\begin{equation}\label{E9}
H \;=\; - \frac{\hbar^2}{2M} \,\left({d^2 \over dr^2}+{1 \over r}{d \over r}-{1
\over
r^2}(m-\alpha)^2\right)
\end{equation}

\noindent where $m=0,1,2,...$ are again the eigenvalues of the angular momentum and
$\alpha=\theta/\pi$. The so(2,1) Lie algebra here is analogous to the one discussed
in the previous section with generators defined by (\ref{T1})-(\ref{T3}). Here the
resolvent operator is simply 
\ba
(H-E)=g_0+g_{1}T_{1}(r)\,,
\ea
with $g_0$ and $g_1$ given by eq. (\ref{gis}) (here $g_3=0$), so that the radial
Green's function can be written as:
\begin{equation}\label{E14}
G(r,r')=\frac i\hbar \int_{0}^{\infty}ds \exp\{-i\frac{s}{\hbar}(g_0-i\epsilon)\}
\exp\{-i\frac{s}{\hbar}g_{1}T_{1}\}{1 \over r}\delta(r-r')
\end{equation}

\noindent It has been shown in ref. \cite{BoschiVaidya90a} that:
\begin{equation}\label{E16}
\exp\{icT_1\}{\delta(r-r') \over r}
=i(-i)^{|m-\alpha|}{M \over s\hbar}
\exp\{{iM \over 2s\hbar}(r^{2}-r'^{2})\}
J_{|m-\alpha|}\left({M \over s\hbar}rr'\right)
\end{equation}

\noindent so that the radial Green's function is given by:
\begin{eqnarray}\label{E17}
G(r,r')=-{M \over \hbar}(-i)^{|m-\alpha|}
\int_{0}^{\infty}{ds \over s}\exp\{i\frac{s}{\hbar}(E+i\epsilon)\}
\exp\{{iM \over 2s\hbar}(r^{2}-r'^{2})\}
J_{|m-\alpha|}\left({M \over s\hbar}rr'\right).
\end{eqnarray}

\noindent Another way to approach the two anyon system without a regulator potential
is to consider the two anyon system in the harmonic well, eq. (\ref{M5A}), and take
the limit where the regulator vanishes $\omega\to 0$. This limit corresponds to take
$k\to 0$ in eq. (\ref{GErr'A}), so that the above Green's function is reobtained. 

This Green's function can be compared with the one obtained for the particle-vortex
system, eq. (\ref{GEm}). One can note that they are identical if one identifies the
quantized flux $\nu$, eq. (\ref{nu}), with the anyon statistical parameter $\alpha$,
eq. (\ref{alpha}).


\section{Two anyons in a uniform magnetic field}


The Hamiltonian of the relative motion of two anyons in a uniform and constant magnetic field $B$ is given by\cite{Arovas:1989ak}-\cite{DasnieresdeVeigy:1993rx}
\begin{equation}\label{mag1}
H=\frac{1}{2\mu}\left(\vec{p}+\frac{1}{2}\mu\omega_{c}r\hat{\phi}+
\frac{\alpha\hbar}{r}\hat{\phi}\right)^2
\end{equation}

\noindent where $\omega_c=eB/mc$ is the cyclotron frequency, the second term on
brackets corresponds to the physical (external) magnetic vector potential $\vec A =
B
r \hat\phi/2$ and the third term is the statistical vector potential. This
statistical term can be absorbed in the angular part of the kinetical term that
contributes to the  angular momentum of the particles. The radial part of this
hamiltonian can be written as before as:
\begin{equation}\label{mag2}
H=-{\hbar^2 \over 2\mu}
\left\lbrack{1 \over r}{\partial \over \partial r}r{\partial \over \partial r}
-{1 \over r^2}\left(m-\alpha\right)^2\right\rbrack
-{m\hbar\omega_c \over 4}+{1 \over 8}\mu\omega_{c}^{2}r^2\mbox{.}
\end{equation}

\noindent The presence of the magnetic field $B$ implies an $r$-independent term
that
contributes to the energy but the form of this Hamiltonian is similar to that of the
problem of two anyons in a harmonic well, discussed in section \ref{Ahw}.
This fact allows the use of the algebraic method as before to calculate the Green's
functions. Then the so(2,1) generators that describe the two anyons in a magnetic
field are the ones given by eqs. (\ref{T1})-(\ref{T3}). The resolvent operator is
again given by eq. (\ref{A1}) with the parameters
\begin{eqnarray}
g_0=-\left(E+\frac{1}{4}m\hbar\omega_{c}\right)\;\mbox{;}
\;\;\;\;
g_1=-{\hbar^2 \over 2\mu}\;\mbox{;}
\;\;\;\;
g_3=-\mu\omega_c^2\;\mbox{.}
\end{eqnarray}

\noindent Following the algebraic method as in section \ref{Ahw} with these
parameters  we find the Green's function:
\begin{eqnarray}\label{mag4}
G_{E}(r,r^\prime,\phi,\phi^\prime)&=&-\frac{1}{\pi}\sum_{m=-\infty}^{+\infty}e^{2\pi
i|m-\alpha|}\left({\mu\omega_c \over 2\hbar}\right)^{1+|m-\alpha|}
(rr^{\prime})^{|m-\alpha|}e^{-im(\phi-\phi')}\nonumber\\
&\times&\sum_{n=0}^{\infty}{n!L_{n}^{|m-\alpha|}
\left({\mu\omega_c \over 2\hbar}r^2\right)L_{n}^{|m-\alpha|}
\left({\mu\omega_c \over 2\hbar}(r^\prime)^2\right) \over
\Gamma(n+|m-\alpha|+1)}\nonumber\\
&\times& \exp \left\{ -{\mu\omega_c \over 4\hbar}(r^2+{r^\prime}^2)\right\}
{1 \over E-{\hbar\omega_c \over 2}(2n +|m-\alpha|+1+ \frac m2)}\,.
\end{eqnarray}

From this Green's function we obtain the normalized wave functions:
\begin{eqnarray}\label{mag5}
\psi_{n,m}^{\alpha}(r,\phi)&=&\frac{i}{\sqrt{\pi}}e^{i\pi|m-\alpha|}e^{-im\phi}
\left({\mu\omega_c \over 2\hbar}\right)^{{1 \over 2}(1+|m-\alpha|)}
r^{|m-\alpha|}\nonumber\\
&\times&{\sqrt{n!}L_{n}^{|m-\alpha|}
\left({\mu\omega_c \over 2\hbar}r^2\right) \over \sqrt{\Gamma(n+|m-\alpha|+1)}}
\exp \left\{ -{\mu\omega_c \over 4\hbar}r^2\right\}\mbox{.}
\end{eqnarray}

\noindent and the corresponding energy levels:
\begin{eqnarray}
E_{nm}^{\alpha}={\hbar\omega_c \over 2}(2n +|m-\alpha|+1+ \frac m2)\;.
\end{eqnarray}
These energy levels coincide with the Landau levels if the anyon statistical
contribution vanishes ($m=\alpha=0$).
In particular, the ground state wave function is obtained when
one takes $m=n=0$:
\begin{eqnarray}\label{mag6}
\psi_{0,0}^{\alpha}(r)=\frac{i}{\sqrt{\pi}}e^{i\pi\alpha}
\left({\mu\omega_c \over 2\hbar}\right)^{{1 \over 2}(1+\alpha)}
{r^{\alpha} \over \sqrt{\Gamma(1+\alpha)}}
\exp \left\{ -{\mu\omega_c \over 4\hbar}r^2\right\}\mbox{.}
\end{eqnarray}


\section{Conclusions}


In this paper we have calculated algebraically the Green's functions for
one and two particles confined on a plane, namely the particle-vortex system and a
pair of anyons with and without external potentials. The external potentials
considered were a harmonic well and a uniform magnetic field. In these
problems we have identified the hamiltonian operator in each case with the generators
of the SO(2,1) Lie group satisfying the so(2,1) Lie algebra. From these algebraic
properties we obtained all relevant dynamical quantities of each system. This means
that these systems are described by the so(2,1) dynamical algebra.

In particular, we calculated the Green's function for particle-vortex
system and the Green's function for a pair of free anyons and found that these
Green's functions are equivalent, once one identifies the quantized flux $\nu$ of the
particle-vortex system with the anyon statistical parameter $\alpha$. 
These two Green's functions exhibit respectively the phase factors $(-i)^{|m-\nu|}$
and  $(-i)^{|m-\alpha|}$ as a commom signature of fractional statistics for both
systems.

We obtained also the Green's function for the two anyon system in a harmonic well as
an integral representation of Bessel functions, and as a sum of product of Laguerre's
generalized polynomials. This sum is recognized as the spectral representation of the
Green's function from which we identify the normalized wave functions and energy
spectrum.

It is interesting to note that in the particle-vortex discussion presented by Jackiw \cite{Jackiw90}, he obtained a discrete spectrum of the generator $R$ (our eq. (\ref{Rjac})) of  conformal transformations, and them, by a rotation in operators space he obtain the continuous spectrum calculated by standard methods to the  particle-vortex system. This discrete spectrum is identical to one obtained for anyons in presence of an harmonic regulator. The eigenfunctions (up to a phase factor) and eigenvalues of this operator, eq. (48) in \cite{Jackiw90}, are equivalent to those ones we have obtained for the two anyon system in a harmonic well. 

Regarding the Green's functions for the two anyon system in a harmonic well and two anyons in a uniform magnetic field both lead to bound states and we see that they are very similar to each other, although they differ in the energy spectrum.  
Charged anyons orbiting in a uniform magnetic field are equivalent to anyon bound states
in the presence of a harmonic well. 
Note that wave functions (\ref{wavef}) and (\ref{mag5}), obtained from those Green's functions, are identical if we identify $\omega$ with $\omega_c/2$.

In particular the ground state wave function obtained for two charged particles
in a magnetic field, eq. (\ref{mag6}), is similar to the two particle wave function used by Laughlin \cite{Laughlin90} to construct his ansatz for N-particles to describe the quantum Hall effect. In the problem discussed by Laughlin, there is also a coulombic interaction which in general could not be disregarded.
Then, he supposed that this interaction is infinitely short ranged and that the 
Landau levels energy is dominant so $\hbar\omega_c >> e^{2}/l$, where
$l$ is the magnetic length. Since the particles are separated by some finite length, this 
allows one to build an ansatz for the many particle ground state wave function as a superposition of single particle wave functions:
\begin{eqnarray}\label{gs}
\psi_{00}^{\alpha}(r_{p,q})=e^{i\pi\alpha}\frac{i}{\sqrt{\pi\Gamma(1+\alpha)}}
\left(\frac{\mu\omega_c}{2\hbar}\right)^{\frac 12(1+\alpha)}
\prod_{p < q}r_{p,q}^{\alpha}\exp{\left(-\frac{\mu\omega_c}{4\hbar}\sum_{p <
q}r_{p,q}^{2}\right)}
\end{eqnarray}
This is essentially the Laughlin's ansatz for N-particles\cite{Laughlin90}. Note that the interchange of particle positions adds a phase to the wave function in agreement with \cite{Forte92}.

Let us now consider the excited states for the two anyon system in a uniform magnetic field. The wave functions (\ref{mag5}), or equivalently (\ref{wavef}), represent the excited states of this system. If we follow a similar reasoning as the ones that support (\ref{gs}), as discussed in \cite{Laughlin83,Laughlin83B,Laughlin90}, 
 we can superpose (\ref{mag5}) to obtain an ansatz for the many particle excited state wave function:
\begin{eqnarray}
\psi_{n,m}^{\alpha}(r,\phi)&=&ie^{i\pi|m-\alpha|}
\frac{\sqrt{n!}}{\sqrt{\pi}\sqrt{\Gamma(n+|m-\alpha|+1}}\left({\mu\omega_c \over
2\hbar}\right)^{{1 \over 2}(1+|m-\alpha|)}\nonumber\\
&\times& e^{-im\phi}
\prod_{i<j}r_{ij}^{|m-\alpha|}
L_{n}^{|m-\alpha|}\left({\mu\omega_c \over 2\hbar}r_{ij}^2\right)
\exp \left\{ -{\mu\omega_c \over 4\hbar}\sum_{i<j}r_{ij}^2\right\}\mbox{.}
\end{eqnarray}
\bigskip

\noindent This wave function is formally in agreement with the result for 
 many anyons obtained by Dunne et all \cite{Dunne:1991bc}.

\bigskip

\noindent {\bf Acknowledgment}: 
H.B.-F. is partially supported by CNPq (Brazilian agency).




\end{document}